\documentstyle[aps,prl,preprint]{revtex}

\begin{document}
\title{Entanglement of individual photon and atomic ensembles }
\author{Guo-Ping Guo\thanks{%
Electronic address: harryguo@mail.ustc.edu.cn } and Guang-Can Guo}
\address{Key Laboratory of Quantum Information, University of Science and Technology\\
of China, Chinese Academy of Science, Hefei, Anhui, P. P. China, 230026}
\maketitle

\begin{abstract}
Here we present an experimentally feasible scheme to entangle flying qubit
(individual photon with polarization modes) and stationary qubit (atomic
ensembles with long-lived collective excitations). This entanglement
integrating two different species can act as a critical element for the
coherent transfer of quantum information between flying and stationary
qubits. The entanglement degree can be also adjusted expediently with linear
optics. Furthermore, the present scheme can be modified to generate this
entanglement in a way event-ready with the employment of a pair of entangled
photons. Then successful preparation can be unambiguously heralded by
coincident between two single-photon detectors. Its application for
individal photons quantum memory is also analyzed. The physical requirements
of all those preparation and applications processing are moderate, and well
fit the present technique.

PACS number(s): 03.65.Ud, 03.67.-a, 42.50.Gy. 42.50.-p
\end{abstract}

\newpage

Quantum entanglement is a fancy correlation between two or more distant
subsystems which has no classical analog. Such a correlation has found wide
applications in high-precision spectroscopy\cite{qs}, quantum lithography%
\cite{ql}, and quantum information processing including computation,
communication and cryptography\cite{qc,qcc}. Up to now individual atoms or
photons have been entangled with a linear ion-trap\cite{it}, with a
spontaneous parametric down converter\cite{spdc1,spdc2}, or with a high-$Q$
cavity\cite{hq,qe}. There is also entanglement between indistinguishable
atoms in Bose-Einstein condensates\cite{bec} or atomic ensembles\cite
{ae1,ae2,dl,g}. In most of those schemes the entangled subsystems are all
congeners (the same kind of objects).

From the beginning of information, light has been believed and used as an
ideal information carrier. It is well known that the photon polarization
modes can be conveniently manipulated with linear optics. Thus individual
photon is a perfect chosen for flying qubit in the nearly developed quantum
information theory. On the other hand, atomic ensemble has enabled it as a
well qualified candidate for stationary and register qubits of quantum
information and computation. Its collective excitation modes have the
inherent robust to realistic noise and imperfections. Quantum repeater based
on atomic ensembles has recently been proposed and shown to be useful in the
long distance quantum communications\cite{dn}. It has been also proved that
the emission of a photon in the forward direction is correlated with an
excitation in a symmetric mode of the atomic ensembles\cite{pra}.

Here we present an idea to entangle two subsystems of different species,
individual photon and atomic ensemble, to integrate the features of both
flying qubit (individual photon) and stationary qubit (atomic ensemble). The
entanglement degree can be also adjusted expediently with linear optics.
This entanglement integrate two different congeners can act as a critical
element for the coherent transfer of quantum information between flying and
stationary qubits. Furthermore, this entangled state can be prepared in a
way event-ready, and successful preparation can be unambiguously heralded by
coincident between two single-photon detectors. As an example of its various
applications, we propose a quantum memory for individual photons based on
this novel entanglement.

A quantum memory for individual photons is obviously valuable in many
quantum information processing such as quantum cryptography\cite
{qc,qcc,qb,or}, and secret sharing\cite{gg}. Distinct from the previous
light memory schemes\cite{jc,qm1,qm2,qm3}, this quantum memory employs a
procedure similar to quantum teleportation and is designed to store
individual photon modes. As the qubit needed store and the memory qubit
don't directly interact with each other, a higher memory fidelity can be
expected in the present quantum memory. As most quantum protocols with
atomic ensembles\cite{ae2,dl,g,dn}, the physical requirements of our
entanglement preparation and applications are moderate and well fit the
present technique.

The basic element of our system is an ensemble of alkali atoms with the
relevant level structure shown as Fig. 1. A pair of metastable low states $%
\left| g\right\rangle $ and $\left| s\right\rangle $ correspond to Zeeman
sublevels of the electronic ground state of alkali-metal atoms. Its
experimental realization can be either a room-temperature atomic gas or a
sample of cold trapped atoms where long lifetimes for the relevant coherence
has been both observed\cite{dn20,d14,d15}. To facilitate enhanced coupling
to light, the atomic medium is preferably optically thick along one
direction. This can be achieved either by working with a pencil-shaped
atomic sample\cite{dn20,d14,d15} or by placing the sample in a low-finesse
ring cavity\cite{dn17,dn25}.

Two identical atomic samples are placed in a beeline with a half-wave plate
and a polarization beam-splitter plate (PBS) between them just as shown in
Fig. 2. Define an operator $S=(1/\sqrt{N_a})\sum_{i=1}^{N_a}$ $\left|
g\right\rangle _i\left\langle s\right| $ where $N_a\gg 1$ is the total atom
number. All atoms are initially prepared through optical pumping to the
ground state $\left| g\right\rangle ,$ which is effectively the vacuum state 
$\left| 0\right\rangle _a=\otimes _i\left| g\right\rangle _i$ of the
operator $S$. Those two atomic samples are illuminated by a short,
off-resonant laser pulse in turn that induces atom Raman transitions into
state $\left| s\right\rangle $.

What particularly interests us is the forward-scattered Stokes light that is 
$co$-propagating with the laser. As such scattering events are uniquely
correlated with the excitations of the symmetric collective atomic mode $S$,
an emission of the single Stokes photon in the forward direction results in
the state of atomic ensemble given by $S^{\dagger }\left| 0\right\rangle
_a=(1/\sqrt{N_a})\sum_{i=1}^{N_a}\left| s\right\rangle _i=\left|
S\right\rangle .$ The excitations in the mode $S$ can be transferred to
optical excitations and then detected by single-photon detectors\cite
{d14,d15,d17}. Due to the collectively enhanced coherent interaction, the
efficiency of such transfer can be very high, which has been demonstrated
both in theory\cite{d17} and in experiments\cite{d14,d15}.

Assume the Stokes photon from this emission is right-handed rotation, and
define an effective single-mode bosonic operator $a_R$ for this Stokes pulse
with the corresponding vacuum state denoted by $\left| 0\right\rangle _p$.
The $\lambda /2$ plate between the two samples is employed to transform this
Stokes photon into left-handed rotation, where $\lambda $ is the wavelength
of the Stokes photon. Its function can be denoted by operator $P=\left|
a_L^{\dagger }\right\rangle \left\langle a_R\right| +\left| a_R^{\dagger
}\right\rangle \left\langle a_L\right| $ with $L$ ($R$) represents the left
(right) handed rotation. We can assume this half-wave plate has no
remarkable influence on the interaction efficiency of the pump light and the
atomic ensemble. The forward-scattered Stokes pulses are collected and
coupled to optical channels (such as fibres) after a filter, which is
frequency selective to filter out the pumping light. In the case there is
only one Stokes photon, the whole system of the two atomic ensembles and
this photon can be written in state 
\begin{equation}
\left| \Phi \right\rangle _{ap}=(\alpha S_1^{\dagger }Pa_R^{\dagger }+\beta
S_2^{\dagger }a_R^{\dagger })\left| 0\right\rangle _a\left| 0\right\rangle
_p,
\end{equation}
where 1 and 2 denotes the two atomic samples respectively. We can assume
that the two atomic samples are identical and the phase difference depending
on the quantum channel between the two samples is fixed as $\alpha =\beta =1/%
\sqrt{2}$. In fact, the two ensemble needn't be identical and only the
parameters $\alpha ,\beta $ are affected. But we can adjust those parameters
to alter the entanglement degree of the above state with the polarization
beam-splitter plate placed between the two atomic ensembles. This
polarization plate controls the pass ratio of the left-handed rotation
Stokes photon from the first atomic ensemble. Finally the Stokes photons
pass through a $\lambda /4$ plate which transforms the circularly-polarized
wave to the linearly polarized light. Here $\lambda $ is again the
wavelength of the Stokes photon. Then $\left| \Phi \right\rangle _{ap}$ can
now be written as the maximally entangled EPR state 
\begin{eqnarray}
\left| \Psi \right\rangle _{ap} &=&(S_1^{\dagger }a_L^{\dagger
}+S_2^{\dagger }a_R^{\dagger })/\sqrt{2}\left| 0\right\rangle _a\left|
0\right\rangle _p  \nonumber \\
&=&(\left| S_1\right\rangle _a\left| H\right\rangle _p+\left|
S_2\right\rangle _a\left| V\right\rangle _p)/\sqrt{2}.
\end{eqnarray}
Here we have assumed that the light-atom interaction time $t_\Delta $ is
short and the mean photon number in the forward-scattered Stokes pulse is
much smaller than 1. The probabilities for more than one photon $p_0^n$ will
be so small that they can be neglected safely, where $n$ represents the
number of the Stokes photons. As we will show below, the including of the
case that there is no Stokes photon scattered out is insignificance.

Noticeably, the subsystems of this entanglement are of different species,
individual photon and atomic ensembles. It is well known that the collective
state of atomic ensemble has inherent resilience to noise and imperfections,
which makes it well fit to act as quantum memory (stationary qubit). On the
other hand, the light is believed to be an ideal carrier (flying qubit) of
quantum information and the polarization states of individual photon can be
conveniently manipulated with linear optics. Although Raman processing to
generate Stokes photons is random, and the probability $p_0$ to prepare the
state$\left| \Phi \right\rangle _{ap}$ is quite low, various applications
can be expected for this novel entanglement combining the features of both
flying and stationary qubits. As most entanglement generation protocols,
postselection may be needed in those applications.

Alternatively, we can employ a quantum non-demolition measurement device\cite
{qnd,g0} to sense the presence of the Stokes photons and wipe off
postselection. This device has been shown implementable with an optical
interferometer\cite{qnd}. However, it has shown recently that one can
conditionally prepare a pair of photons in maximally entangled polarization
state when smaller coefficient higher terms are negelected\cite{qqq}. The
successful preparation can be unambiguously heralded. With a pair of
entangled photons, we can further modify the present protocol to generate
entanglement state between flying and stationary qubits in an event-ready
way.

Assume we have a pair of photons in state $\left| \Phi \right\rangle
_{AB}^{+}=$ $(\left| HH\right\rangle _{AB}+\left| VV\right\rangle _{AB})/%
\sqrt{2}$. As shown in Fig. 2, we can incident the Stokes photons generating
from the atomic ensembles, and the photons $A$ simultaneously onto a
beam-splitter plate. The whole system can then be written in state

\begin{equation}
\left| \Psi \right\rangle _{apAB}=(\left| 00\right\rangle
_{ap}+p_0^{1/2}\left| \Phi \right\rangle _{ap}+o(p_0))\otimes \left| \Phi
\right\rangle _{AB}^{+},
\end{equation}
where $o(p_0)$ represents the term involving more than one Stokes photon. In
the expanding of this state, only the second term with a coefficient $%
p_0^{1/2}$ totally involve three photons: 
\begin{eqnarray*}
&&\left| \Phi \right\rangle _{ap}\otimes \left| \Phi \right\rangle _{AB}^{+}
\\
&=&(\left| S_1\right\rangle _a\left| H\right\rangle _p+\left|
S_2\right\rangle _a\left| V\right\rangle _p)/\sqrt{2}\otimes (\left|
HH\right\rangle _{AB}+\left| VV\right\rangle _{AB})/\sqrt{2} \\
&=&\frac 12\{\left| \Phi \right\rangle _{pA}^{+}(\left| H\right\rangle
_B\left| S_1\right\rangle _a+\left| V\right\rangle _B\left| S_2\right\rangle
_a)/\sqrt{2}+\left| \Phi \right\rangle _{pA}^{-}(\left| H\right\rangle
_B\left| S_1\right\rangle _a-\left| V\right\rangle _B\left| S_2\right\rangle
_a)/\sqrt{2} \\
&&+\left| \Psi \right\rangle _{pA}^{+}(\left| V\right\rangle _B\left|
S_1\right\rangle _a+\left| H\right\rangle _B\left| S_2\right\rangle _a)/%
\sqrt{2}+\left| \Psi \right\rangle _{pA}^{-}(\left| V\right\rangle _B\left|
S_1\right\rangle _a-\left| H\right\rangle _B\left| S_2\right\rangle _a)/%
\sqrt{2}\},
\end{eqnarray*}
where $\left| \Phi \right\rangle _{pA}^{\pm }=$ $(\left| HH\right\rangle
_{pA}\pm \left| VV\right\rangle _{pA})/\sqrt{2},\left| \Psi \right\rangle
_{pA}^{\pm }=$ $(\left| HV\right\rangle _{pA}\pm \left| VH\right\rangle
_{pA})/\sqrt{2}$ are the four Bell states. When there are coincidence clicks
between single-photon detectors placed on each side after the beam-splitter, 
$D_H$ and $D_{V^{\prime }}$ or $D_V$ and $D_{H^{\prime }}$, the two photons
are measured in state$\left| \Psi \right\rangle _{pA}^{-}$. In this case the
two atomic samples and photon $B$ are projected into the state $\left| \Psi
\right\rangle _{aB}^{-}=(\left| V\right\rangle _B\left| S_1\right\rangle
_a-\left| H\right\rangle _B\left| S_2\right\rangle _a)/\sqrt{2}$. Similarly,
if there are coincidence clicks between single-photon detectors behind the
two-channel polarizer (PBS), $D_H$ and $D_V$ or $D_{V^{\prime }}$ and $%
D_{H^{\prime }}$, then two photons $A$ and $p$ are measured in state $\left|
\Psi \right\rangle _{pA}^{+}$. And the atomic samples are projected into
state $\left| \Psi \right\rangle _{aB}^{+}=(\left| V\right\rangle _B\left|
S_1\right\rangle _a+\left| H\right\rangle _B\left| S_2\right\rangle _a)/%
\sqrt{2}$. Obviously, this state can be simply transformed into the state $%
\left| \Psi \right\rangle _{aB}^{-}$. Thus we totally have a probability of $%
p_0/2$ to generate a maximally entangled state between the two ensembles and
one photon. And the successful preparation is unambiguously heralded by
coincidence detection of two photons.

As no postselection is needed, this event-ready entangled state between
stationary and flying qubits can be used straightforwardly used in various
quantum information processing. For example, this entangled state can be
explored as a valuable individual photons quantum memory. This can be
accomplished just by measuring the photon needed storing $q$ in state $%
\left| \Phi \right\rangle _q=\cos \theta \left| H\right\rangle _q+\sin
\theta \left| V\right\rangle _q$ and the photon $B$ of the state $\left|
\Psi \right\rangle _{aB}^{-}$, with the same Bell analyzer as shown in Fig.
2. Then the state of the photon $q$ can be transferred to the two atomic
ensembles. The successful probability is $1/2$ as we can only distinguish
two Bell states out of four with linear optics.

With the present technique when it transferred to optical excitations\cite
{d14,d15,d17} we can conveniently readout the quantum states stored in
atomic ensembles. Remarkably quantum state exchange in this quantum memory
is between individual photon and atomic ensembles. This exchange is an
important ingredient for the future quantum information networks\cite{jc}.
It is also crucial for sensitive atomic measurements in optics when quantum
limits of accuracy are approached\cite{qs}. Distinct from the previous
quantum memory\cite{jc,qm1,qm2,qm3}, the present quantum memory has
following features: firstly, the polarization states of individual photon is
stored, which is used to encode information in many quantum computation and
information proceeding. Secondly, the qubit needed storing doesn't interact
with the memory qubit. It is in based on a memory procedure similar to
quantum teleportation. Thirdly, there is no complex parameters as
interaction time and conditions to effect the memory fidelity. Improving
Bell measurement efficiency of the quantum teleportation procedure, we can
enhance the memory fidelity which is defined as $\left| \left\langle \Phi
\right| \Psi \rangle \right| ^2$ with $\left| \Phi \right\rangle $ and $%
\left| \Phi \right\rangle $ represent the states before and after storing.
Finally, the physical requirements of the individual photon memory scheme
are moderate and well fit the experimental technique.

Finally, we briefly discuss the matters of the experimental implication. As
the light-atom interaction time $t_\Delta $ is short and the mean photon
number in the forward-scattered Stokes pulse is much smaller than 1, the
whole system of the Stokes photon and atomic ensembles is in the mixed 
\begin{equation}
\rho _{ap}=\frac 1{p_0+1}(\left| 00\right\rangle _{ap}\left\langle 00\right|
+p_0\left| \Psi \right\rangle _{ap}\left\langle \Psi \right| ).
\end{equation}
This mixed state will be purified automatically to the maximally entangled
state in the entanglement-based communication schemes with the one Stokes
photon coefficient $p_0$ determining the purification efficiency\cite{dn}.
Then this mixed state can be called as an effective maximally entangled
state of individual photon and atomic ensembles. Furthermore, it has been
shown that this scheme is capable of event-ready generation of entanglement
between stationary and flying qubits with a probability of $p_0/2$. In order
to efficiently filter out the pump light from the Stokes photon with the
frequency-selective filter, the energy difference between the ground state
and the excited should be large enough. Similar to the papers\cite{dn,dl},
the single-bit rotation error (below $10^{-4}$ with the use of accurate
polarization techniques for Zeeman sublevels) and the dark count
probability( about $10^{-5}$in a typical detection time window $0.1\mu s$)
of single-photon detectors can be both neglected.

Actually, similar entanglement states can be prepared by using only one
atomic ensemble with the relevant level structure as shown in Fig. 3. The
difficult is the control of the probability amplitudes for the emitted of
two kinds of polarized Stokes photons, which may be much easier in the above
two separated atomic ensembles case. Supplied with multi-photon
entanglement, one can also prepare event-ready multi-party entanglement of
flying and stationary qubits in a similar teleportation way as above.

In summary, we have proposed an experimentally feasible scheme to generate
entanglement between flying qubit (individual photon with polarization
modes) and stationary qubit (atomic ensembles with long-lived collective
excitations). This entanglement integrate two different species can act as a
critical element for the coherent transfer of quantum information between
flying and stationary qubits. Furthermore, the present scheme is modified to
generate this entanglement between stationary and flying qubits in a way
event-ready. Successful preparation can be unambiguously heralded by
coincident between two single-photon detectors. The entanglement degree can
be also adjusted expediently with linear optics. As an example of its
various applications, we propose a high fidelity quantum memory for
individual photons, which is based on a procedure similar to quantum
teleportation. As those protocols based on atomic ensembles, the physical
requirements of all the preparation and applications of the present scheme
are moderate and well fit the present technique. Their experimental
implementations can be expectable in near future.

This work was funded by National Fundamental Research Program(2001CB309300),
National Natural Science Foundation of China, the Innovation funds from
Chinese Academy of Sciences, and also by the outstanding Ph. D thesis award
and the CAS's talented scientist award entitled to Luming Duan.

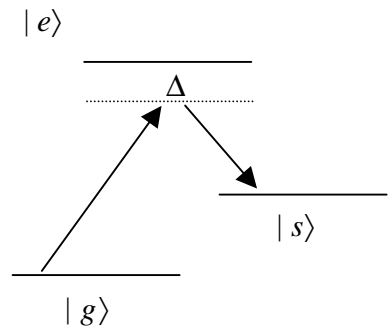
Figure 1: The relevant level structure with $\left| g\right\rangle ,$ the
ground state, $\left| e\right\rangle ,$ the excited state, and $\left|
s\right\rangle $ the metastable state for storing a qubit. The transition $%
\left| g\right\rangle \rightarrow \left| e\right\rangle $ is coupled by the
classical laser (the pump light) with Rabi frequency $\Omega ,$ and the
forward-scattered Stokes light comes from the transition $\left|
e\right\rangle \rightarrow \left| s\right\rangle ,$ which is right-handed
rotation. For convenience, we assume off-resonant coupling with a large
detuning $\Delta .$

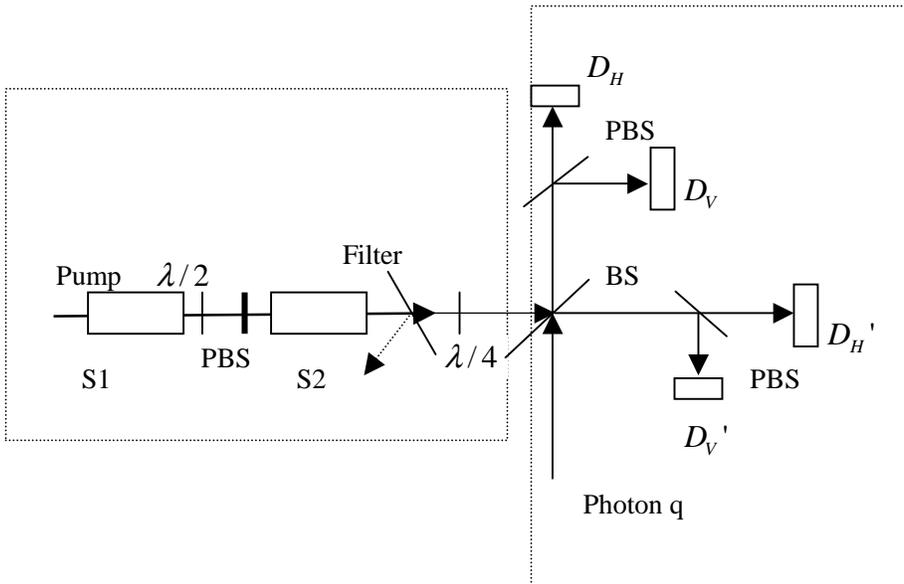
Figure 2: Schematic setup-up for the preparation and application for quantum
memory of entanglement between the Stokes photon $p$ and the two atomic
ensembles $S_1$ and $S_2$. The $\lambda /2$ plate transforms the Stokes
photon scattered out from the sample $S_1$ into the left-handed rotation.
The polarization beam splitter plate (PBS) between the two atomic ensembles
is employed to adjust the entanglement degree. The frequency-selective
filter separate the pump light from the Stokes photon. And the $\lambda /4$
transforms the circularly-polarized wave to the linearly polarized light.
This Stokes photon $p$ and the addition photon $A$ or $q$ interfere at a $%
50\%-50\%$ beam splitter BS, with the outputs analyzed by two double-channel
polarisers respectively and detected by four single-photon detectors $D_H$, $%
D_{V^{\prime }}$, $D_V$ and $D_{H^{\prime }}.$ In fact this a Bell-state
analyzer. The coincidence clicks between $D_H$ and $D_{V^{\prime }}$ or $D_V$
corresponds $\left| \Psi \right\rangle _{pq}^{-}$ and the coincidence clicks
between $D_H$ and $D_V$ or $D_{V^{\prime }}$ and $D_{H^{\prime }}$
corresponds $\left| \Psi \right\rangle _{pq}^{+}$.

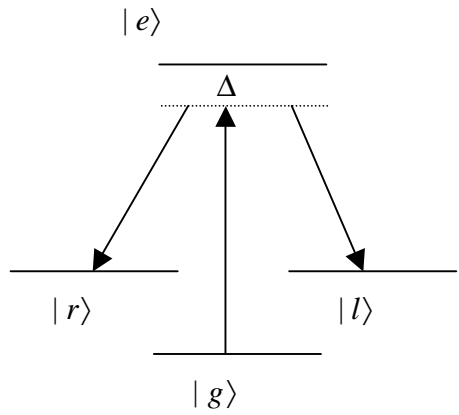
Figure 3: The relevant level structure with $\left| g\right\rangle ,$ the
ground state, $\left| e\right\rangle ,$ the excited state, and $\left|
r\right\rangle $ , $\left| l\right\rangle $ the two metastable state for
storing a qubit. The transition $\left| g\right\rangle \rightarrow \left|
e\right\rangle $ is also coupled by the classical laser (the pump light)
with Rabi frequency $\Omega ^{\prime },$ and the two kinds of
forward-scattered Stokes light comes from the transition $\left|
e\right\rangle \rightarrow \left| r\right\rangle $ and $\left|
e\right\rangle \rightarrow \left| l\right\rangle $ which are right-handed
and left-handed rotation respectively. For convenience, we assume
off-resonant coupling with a large detuning $\Delta .$

\end{document}